\documentclass[useAMS,amssymb,epsfig,usegraphicx,twocolumn,usenatbib]{mn2e}
\def\aj{AJ}

\def\apj{ApJ}
\def\apjl{ApJL}
\def\apjs{ApJS}
\def\aap{A\&A}
\def\mnras{MNRAS}
\def\na{NewA}

\usepackage{graphicx,times}

\def\deg{$^{\circ}$}

\def\kpc{{\rm kpc}}

\usepackage[usenames,dvips]{color}

\title[Formation of double bars ]
{Spontaneous formation of double bars in dark matter dominated galaxies}

\author[Saha et al.]
{Kanak Saha$^{1,2}$\thanks{E-mail:saha@mpe.mpg.de}, \& Witold Maciejewski$^{3}$\\
1. Max-Planck-Institut f\"ur Extraterrestrische Physik, Giessenbachstrasse, D-85748 Garching, Germany\\
2. Inter-University Centre for Astronomy and Astrophysics, Post Bag 4, Ganeshkhind, Pune 411007, India\\
3.Astrophysics Research Institute, Liverpool John Moores University, Twelve Quays House, Egerton Wharf, Birkenhead CH41 1LD, UK\\}
 
\begin{document}

\date{Accepted xxxx Month xx. Received xxxx Month xx; in original form
2012 Nov. 29} 
\maketitle

\label{firstpage}

\begin{abstract}
Although nearly one-third of barred galaxies host an inner, secondary
bar, the formation and evolution of double barred galaxies remain unclear.
We show here an example model of a galaxy, dominated by a live dark matter 
halo, in which double bars form naturally, without requiring gas, and we 
follow its evolution for a Hubble time. The inner bar in our model galaxy 
rotates almost as slowly as the outer bar, and it can reach up to half of
its length. The route to the formation of a double bar may be different 
from that of a single strong bar. Massive dark matter halo or dynamically hot stellar
disc may play an important role in the formation of double bars and 
their subsequent evolution.
\end{abstract}


\section{Introduction}
\label{sec:introduc}
A high fraction ($> 60\%$) of disc galaxies in the local Universe 
are barred, including our Milky Way, 
and nearly 30\% of barred galaxies host an inner, secondary bar,
nested inside the main bar, i.e., are double barred
\citep{ErwinSprake2002,Erwin2011,Laineetal2002}. 
The inner bars are likely to be old structures, as they are seen in the 
near-infrared \citep{Mulchaeyetal1997} and stellar population analysis
gives rather old age estimates \citep{deLorenzoetal2012,deLorenzoetal2013}.
There is observational evidence that the two bars in double barred 
systems rotate independently \citep{Corsinietal2003}.
The observed common occurrence of double bars is not reflected in the present
numerical models of bar formation. While a stellar disc with the Toomre parameter
$Q\sim 1$ readily forms a single bar, which then grows within a Gyr to become a strong 
bar (such as the one in NGC 1300), long-term evolution of such a strong bar 
does not lead to the formation of a double bar on its own in purely stellar 
discs. Only a few N-body simulations have reported formation of double bars 
\citep{RautiainenSalo1999, Curiretal2006, DebattistaShen2007}, mostly when 
special initial conditions have been imposed.

In $2D$ simulations of \cite{RautiainenSalo1999}, with rigid bulge and halo, 
double bars form when the Toomre $Q$ parameter is increased in the 
central parts of the disc, and they can survive for the Hubble time.
In sets of cosmological $N$-body models by \cite{Curiretal2006},
double bars form when the mass of the disc is lowered.
\cite{DebattistaShen2007} showed that the inner bar developing from a 
rapidly rotating bulge, or a pseudobulge, survives many relative rotations of
the bars. Such inner bar pulsates, and its pattern speed oscillates 
in accord with predictions from orbital analysis \citep{MaciejewskiSparke1997,MS2000}.
Another route to forming double bars relies on the presence of a dissipative 
(gaseous) component. In this scenario \cite[e.g.,][]{FriedliMartinet1993}, 
gas inflow in the large-scale bar stagnates in the inner kpc, leading to the 
formation of a disc there, which may become unstable and give rise to a 
smaller, secondary bar. However, in numerical realizations of this scenario, 
the inner bar lasts no longer than a few relative rotations of the bars 
\cite[see sect.5.2 of][for a summary]{MA2008}.

In this letter, we report spontaneous formation of double bars in a 
dark matter dominated stellar disc without any gas.
We have performed a suite of simulations of dark-matter-dominated galaxies 
(Saha et al. in prep) in 3 dimensions that include a live halo. We noticed 
that in a few cases, structures resembling double bars form naturally 
in our simulations. In this paper, we present a model with such structure
being most evident. It provides insight into factors decisive in formation
of double bars, and explores self-consistent double bars of parameters
markedly different from previous simulations.

\section{Initial galaxy model}
\label{sec:modelsetup}
Equilibrium model of a galaxy is constructed using the self-consistent method of
\citet{KD1995}. The initial galaxy model consists of a live disc, halo and a classical
bulge. The disc has an exponentially declining surface density with a
scale-length $R_d$, scale-height $h_z$, and mass $M_d$. In internal units, where
G=1, these parameters take the following values: $R_d=1$, $h_z=0.03$ and 
$M_d=1.58$. 
The outer radius of the disc
is truncated at $6.0 R_d$ with a truncation width of $0.3 R_d$ within which the
stellar density smoothly drops to zero. The live dark
matter halo is modelled with a lowered Evans model \citep{Evans1993} which has a
constant density core. Such a cored halo is known to better represent the
observed high resolution rotation curves in low surface brightness (LSB) galaxies
\citep{KuziodeNarayetal2006}.
The initial classical bulge is modelled with a King model 
\citep{King1966}. The mass of the dark halo is $M_h=20.43$ and that of the 
classical bulge is $M_b = 0.153$. For relevant details on model
construction, the reader is referred to \cite{Sahaetal2010, Sahaetal2012}. 
The initial Toomre $Q$ profile for the
galaxy is such that $Q$ rises to a high value beyond about $5 R_d$ and the same
happens at radius below $\sim 1 R_d$. At $2.5 R_d$, the Toomre $Q$ reaches the
minimum of $Q= 2.55$. 

\begin{figure}
\rotatebox{270}{\includegraphics[height=7.5 cm]{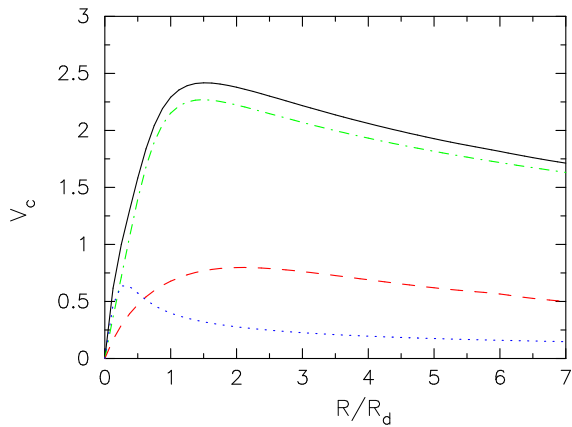}}
\caption{Initial circular velocity. Red line denotes contribution from the 
disc, blue and green lines denote contributions from the bulge and the dark 
matter halo, respectively. Solid black line is the total circular velocity. 
Length and velocity are given in internal units. See section~\ref{sec:modelsetup} 
for scaling to physical units.}
\label{fig:vc102}
\end{figure}

In Fig.~\ref{fig:vc102}, we show the circular velocity curve for the galaxy model
under consideration. The model galaxy is dark matter dominated right from the
central region. This is a norm amongst most LSB galaxies \citep{deBloketal2001}.
However, rotation curves in LSB galaxies usually show a slow rise, while the 
circular velocity curve in our model rises sharply in the inner region. Such
a sharp rise is seen in giant LSB galaxies which often contain a bulge-like component 
\citep{Beijersbergenetal1999,Lellietal2010}. Our galaxy model has some resemblance
to these giant LSBs, but unlike them it contains no gas; hence caution is advised 
if our results are to be used in studies of LSB galaxies.
If we set the unit of length 
to $R_d=4.0$~kpc and the circular velocity at $R=2.1 R_d$ to $220$~km~s$^{-1}$,
then the units of time, mass and velocity are $42$~Myr, 
$8.08\times10^9$~M$_{\odot}$, and $93.2$~km~s$^{-1}$, respectively. 
Dimensional values in the remainder of this paper are given in this standard scaling.
Thus, in our standard scaling, the disc, bulge and halo masses are
$M_d = 1.27 \times 10^{10} M_{\odot}$, $M_b = 0.124 \times 10^{10} M_{\odot}$,
and $M_h = 1.65 \times 10^{11} M_{\odot}$, respectively. 
For any other mass unit $M_0$ and length unit $R_d$, the time unit is 
$42$~Myr$(R_d/4 \kpc)^{1.5}(M_0/8.08\times10^9$~M$_{\odot})^{-0.5}$
and the velocity unit is 
$93.2$~km~s$^{-1} (R_d/4 \kpc)^{-0.5}(M_0/8.08\times10^9$~M$_{\odot})^{0.5}$.
Note that scale lengths of giant LSB discs are typically 
$10$~kpc or more \citep{Beijersbergenetal1999}, but for such scaling the
simulated evolution time exceeds Hubble time.

\begin{figure*}
\rotatebox{0}{\includegraphics[height=6.5cm]{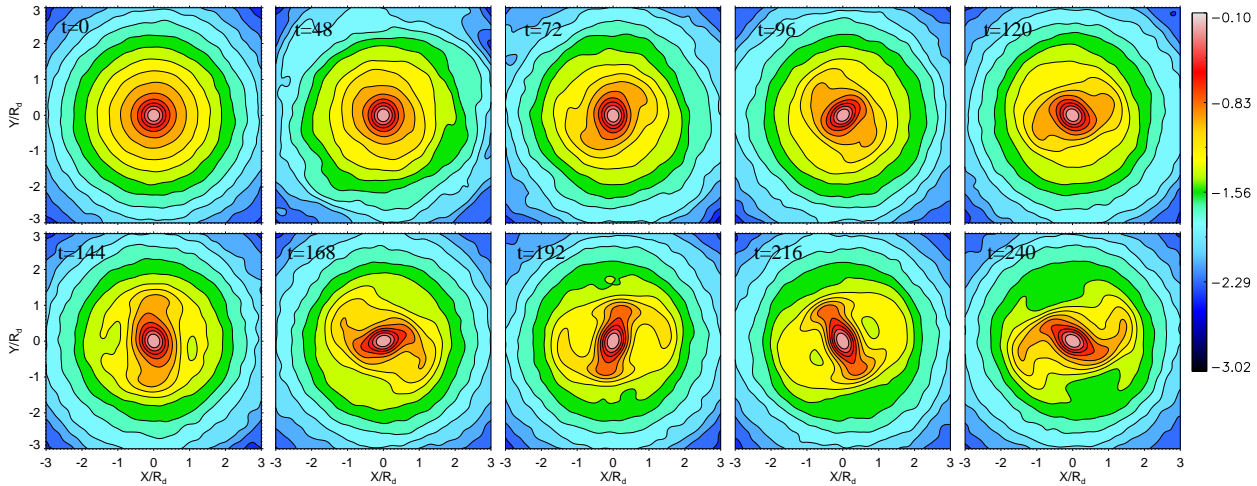}}
\caption{Snapshots of face-on column density of all stars in our galaxy model. 
The color bar represents density in logarithmic scale in internal units, 
which correspond to $505$~M$_{\odot}$~pc$^{-2}$ in our standard scaling. 
The disc rotates counterclockwise.
Snapshot time in internal units is shown in top-left corner of each panel.
The first panel shows the initially axisymmetric disc that subsequently evolves
into a double barred system.}
\label{fig:denmapDB}
\end{figure*}

The simulation was performed using the Gadget code
\citep{Springeletal2001} with a tolerance parameter $\theta_{tol} =0.7$,
and the maximum value of the integration time step $\sim 0.03$, corresponding
to $1.2$~Myr. A total of $2.2$ Million
particles were used to represent the galaxy model with $1.05$ Million each for the disk
and halo and $0.1$ Million for the bulge.

\section{Results}
\label{sec:result}
The set of images presented in Fig.~\ref{fig:denmapDB} represents density 
of the stellar component at different times throughout the run, projected 
onto the disc plane. The radial variation 
of the $m=2$ Fourier component of the stellar density as a function of time
is presented in Fig.~\ref{fig:A2rt}. In order to estimate the extent of 
the bars, and to measure their orientation, we fitted ellipses to the density 
field on a set of images such as in Fig.~\ref{fig:denmapDB}, and derived their 
ellipticity and the position angle (PA) in the same way as it is done for the
observational data.
In Fig.~\ref{fig:ellipiraf}, we show the ellipticity and the PA as a function 
of the semi-major axis (SMA) obtained using the IRAF ELLIPSE fitting routine.
If a bar is present, the PA of the major 
axis should be nearly constant over a range of sizes, with ellipticity 
reaching local maximum within this range. By matching the radial variation 
of the PA and the peak in the ellipticity, we assign an average 
value of the PA to each bar with an average error of $\sim 10$\deg.
When the inner bar is not exactly perpendicular to the outer one, spiral 
features start from the end of the inner bar, making the measurement of the 
PA of the bar difficult. In this situation, we recheck our automated 
measurement of assigning a PA by eye. In measuring the length of the bar, 
we follow the algorithm described by \cite{Erwin2005} for deriving $L_{bar}$ there. 
This is the upper limit for the length of a bar.

\subsection{Formation of two bars}
\label{sec:formation}
\begin{figure}
\rotatebox{0}{\includegraphics[height=4.6 cm]{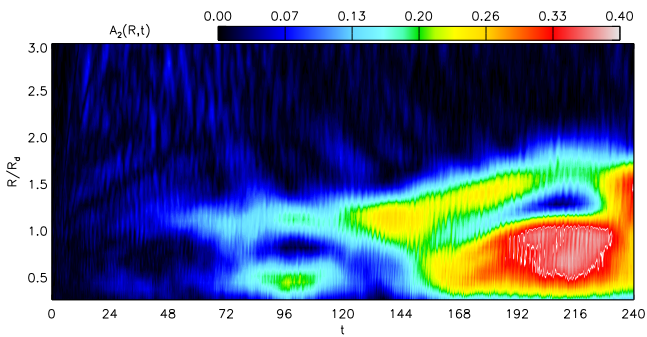}}
\caption{Time evolution of $m=2$ Fourier component ($A_2$) of the stellar 
density distribution as a function of radius in polar coordinates aligned
with the stellar disc. Color bar represents $A_2$ normalized by the 
axisymmetric component $A_0$. The two bars are approximately confined within 
$2$ scale lengths.}
\label{fig:A2rt}
\end{figure}

\begin{figure}
\rotatebox{0}{\includegraphics[height=8.5cm]{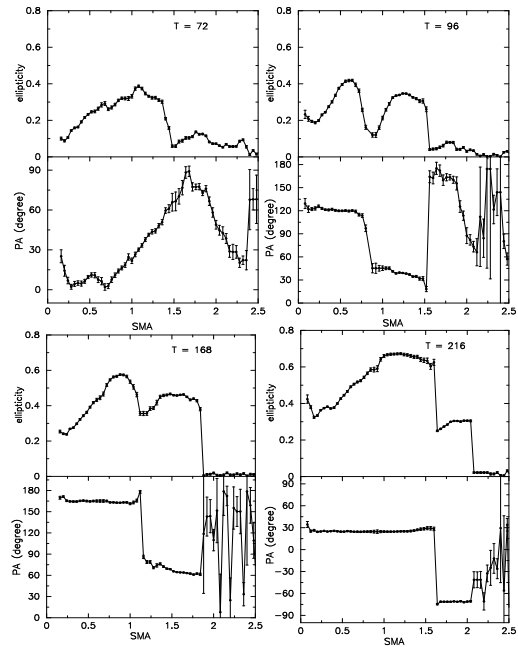}}
\caption{Ellipticity and position angle of ellipses fitted at four instants
(times shown in each panel) to the density distribution from Fig.2 using IRAF 
ELLIPSE routine, as a function of the SMA length. At $t=72$ the bars are not 
well defined yet, at $t=96$ and $t=216$ the bars are almost perpendicular to
each other, and at $t=168$ the bars are out of alignment.}
\label{fig:ellipiraf}
\end{figure}

We follow the evolution of an initially axisymmetric stellar disc embedded 
in a dark matter halo, which gravitationally dominates the stellar component 
throughout the extent of the disc. Since the stellar disc is initially hot, it does not 
form a bar readily. There is no clear non-axisymmetric structure in the disc till t=$48$ in
Fig.~\ref{fig:denmapDB}, which corresponds to $2$~Gyr, but a short open spiral 
can be noticed at t=$72$ ($3$~Gyr) in Fig.~\ref{fig:denmapDB}.
The PA of the fitted ellipses at this time, shown in Fig.~\ref{fig:ellipiraf}, 
increases almost monotonically with radius, hence the two bars are not well defined yet. 
However, there are two local maxima in ellipticity with values higher than 
$0.2$, and the m=2 Fourier component in Fig.~\ref{fig:A2rt} also shows two maxima at $t=72$.
These two maxima, albeit with much lower amplitude, can be traced back to 
$t=48$ at least, with the maximum
corresponding to the outer bar forming first. Thus in our model two 
independent structural components are present from early stages of the run, 
which then develop into two bars. At later times, the two maxima in Fig.~\ref{fig:A2rt}
correspond to the two bars. 

As the asymmetry grows in strength, the spiral transforms into two well 
defined bar-like structures that appear almost simultaneously over 
time between $t=72$ and $96$ ($3$ and $4$~Gyr). At $t=96$,
the two bars are nearly perpendicular to each other. Although there is still 
a spiral transition between the bars, the ellipse PA shown in Fig.~\ref{fig:ellipiraf} 
is roughly constant within the inner (up to $0.75$ SMA length) and the 
outer bar (up to $1.5$ SMA length). The outer bar has a clear boxy isophote.

\begin{figure}
\rotatebox{270}{\includegraphics[height=7.5 cm]{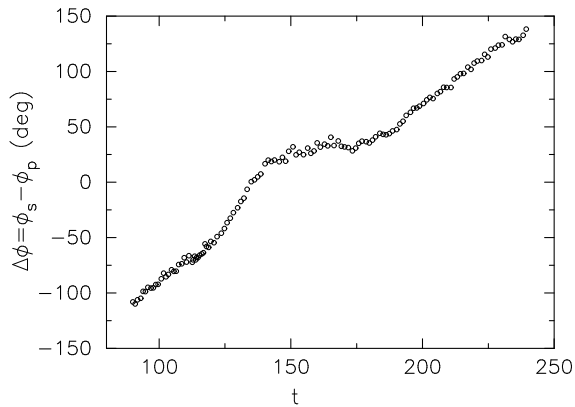}}
\caption{Phase difference (in degrees) between the two bars, 
$\phi_{S} -\phi_{P}$, as a function of time.}
\label{fig:delphi}
\end{figure}

\subsection{Rotation of two bars}
\label{sec:evolution}
After $t=48$, the snapshots in Fig.~\ref{fig:denmapDB} are shown every $24$ time units 
(corresponding to $1$~Gyr), which is approximately two rotation periods 
of the outer bar (see below). The sequence clearly demonstrates that the 
inner bar rotates with respect to the outer bar. However, we find that the 
inner bar definitely is a very slowly rotating structure: it takes several 
rotations of the outer bar for the relative angle between the bars to change 
considerably. During the period of $5$~Gyr (between $t=96$ and $t=216$), the inner bar 
has rotated only once inside the outer bar, going 
from one state when the two bars are orthogonal to the next one. 

\begin{figure}
\rotatebox{270}{\includegraphics[height=7.5 cm]{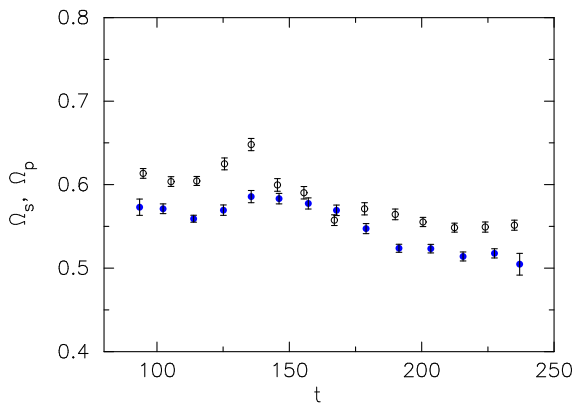}}
\caption{Pattern speed of the two bars as a function of time. Filled blue 
circles denote the primary bar and open black circles correspond to the 
secondary bar.}
\label{fig:pattspd}
\end{figure}

In order to quantify the rotation of the two bars, the PA of the outer bar, 
$\Phi_P$, and of the inner bar, $\Phi_S$, were calculated every $0.3$ time units 
in the inertial frame. During every such interval, the PA of each bar increases
by about 10\deg. In Fig.~\ref{fig:delphi}, we show how the phase difference of 
the two bars, $\Phi_S-\Phi_P$, evolves with time. This difference
increases monotonically, which means that the inner bar rotates faster than
the outer bar. Past $t=190$, the difference becomes linear with time.
In Fig.~\ref{fig:pattspd}, we plot the pattern speed of each bar as a function of time. 
They are derived by fitting consecutive straight lines to $\Phi_P(t)$ and $\Phi_S(t)$
data points over every period when each bar rotates by $360$\deg\ in the inertial frame. 
These measurements are sufficiently accurate to imply that changes in pattern speed 
over time are real, though there is no clear regularity in these changes. The pattern 
speeds of the two bars are different only by $\sim 10$\% at the most. After $t=190$,
the rotation period of the outer bar is $T_P=2\pi/\Omega_P \simeq 12$ time units.  

Although past $t=96$ the two bars are separate entities, as demonstrated by
nearly constant PA of fitted ellipses in Fig.~\ref{fig:ellipiraf}, there is a 
spiral structure between them visible in Fig.~\ref{fig:denmapDB}. When the 
two bars are moving away from alignment 
at $t=136$, towards becoming perpendicular at $t=216$,
this spiral is trailing. It turns to a leading spiral as 
soon as the bars get past the perpendicular arrangement at $t=216$, and are on
their way to become parallel again. Thus the spiral is 
always trailing when the bars are getting out of alignment, and leading when 
they are getting back to alignment. The spiral structures are nearly absent 
when the two bars are perpendicular to each other. These characteristics are
different from a spiral emerging at the ends of a bar that is driven by 
that bar, as in that case the spiral should always be trailing. 
The spiral in our model may be caused by the orientation of orbits in the potential
of the two bars, like in model02 of \cite{Maciejewski-Small2010}, when the
loops (maps of orbits) form a trailing spiral when the bars are leaving
the alignment, and a leading spiral when they are coming back to the alignment.
When the bars are parallel or perpendicular, the loops are aligned, and 
therefore they do not form a spiral shape. The
spiral structure may influence the dynamics of the two
bars, and may indicate that the two bars are dynamically coupled,
although in a different way than having resonances overlapping or pattern
speeds commensurate.

\begin{figure}
\rotatebox{270}{\includegraphics[height=7.0 cm]{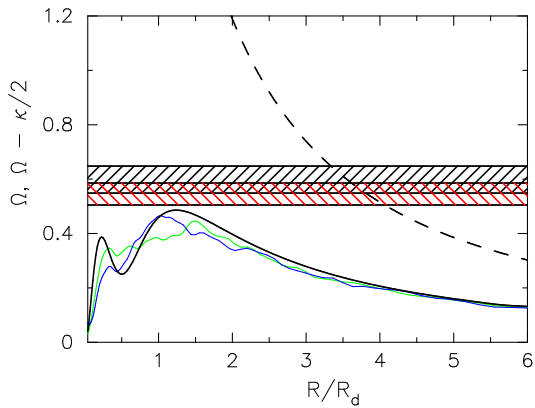}}
\caption{Resonances in the stellar disc. Black solid and dashed curves are 
axisymmetric approximations of $\Omega -\kappa/2$ and $\Omega$, respectively. Blue 
and green curves show $\Omega -\kappa/2$ derived from tangential velocity along the
outer and inner bar, respectively, when the bars are perpendicular ($t=96$). The 
red and black hatched regions denote ranges of pattern speeds of the primary and the
secondary bar, respectively, measured over the time period between t=90 and 
t=240.}
\label{fig:resonan}
\end{figure}

\subsection{Evolution and dynamics of two bars}
The strength and size of the two bars increase with time, as can be seen in
Fig.~\ref{fig:A2rt} and Fig.~\ref{fig:ellipiraf}. Both of these quantities 
can be reliably measured when the two bars are orthogonal to each other, i.e.
at $t=96$ and $216$. Between these two times, the length of the inner bar 
increases by two-fold: from $0.75 R_d$ to $1.55 R_d$ (see Fig.~\ref{fig:ellipiraf}). 
In the same time period, the length of the outer bar grows monotonically from 
$1.5 R_d$ to $2.1 R_d$. Thus the length ratio of the bars appears to increase
from $0.5$ at t=96 to $0.75$ at $t=216$. However, at relative bar positions other 
than orthogonal, the estimate of the length of the inner bar returns lower values, 
possibly because of the presence of a spiral structure connecting the bars, and
then the length ratio remains close to $0.5$.
The amplitude of the peak in ellipticity associated with 
the inner bar grows from $0.4$ at $t=96$ to almost $0.7$ at $t=216$  
(see Fig.~\ref{fig:ellipiraf}), i.e., by a factor 
of $\sim 1.5$. The ellipticity of the outer bar is lower than of the inner one:
at $t=96$ and $216$ it is about $0.3$, although it reaches $0.5$ at $t=168$ and $240$.
The increase of strength and size of the bars is moderated by the relative 
position of the bars: $A_2$ within the outer bar is reduced when the bars are 
orthogonal, and $A_2$ within the inner bar is reduced when the bars are parallel
(Fig.~\ref{fig:A2rt}). The inner bar is growing stronger particularly after 
$t=144$, as it grows in size and its $A_2$ increases. This is because of the 
combination of the secular and the periodic change caused by moving away from alignment.

Having confirmed that the two bars are independent structures, one 
would like to know their dynamics: are they slow or fast bars, and
what resonances they generate. Pattern speed of the outer bar
decreases from about $\Omega_P=0.58$ at $t=136$ to $\Omega_P=0.50$
at $t=237$. Comparing these values with the azimuthal frequency curve in 
Fig.~\ref{fig:resonan}, we have the corresponding corotation radii at 
$3.6$ and $4.1 R_d$. On the other hand, the length of the outer bar increases 
in the same time interval from $1.66$ to $2.25 R_d$. Thus the outer bar is slow, in 
the sense that it extends to only about $0.5$ of its corotation 
radius, with this ratio increasing from $0.46$ to $0.55$. As the pattern
speed of the inner bar is similar to that of the outer bar, the 
inner bar extends to even lower fraction of its corotation radius,
with the ratio around $0.3$.

In order to determine the presence of the Inner Lindblad Resonance (ILR), in 
Fig.~\ref{fig:resonan}, we plot the axisymmetric approximation to the 
$\Omega - \kappa/2$ curve derived from the rotation velocity, accompanied
by the same curves derived from tangential velocity on the major axis
of each bar, which relaxes the assumption of axial symmetry. These curves
do not differ much, which is expected in a model dominated by nearly
spherical dark matter halo. Over the $\Omega - \kappa/2$ curves, we overplot 
the range of pattern speeds associated with each bar throughout the run. 
Our measurements are consistent 
with either no ILR or a weak single ILR at around $1 - 1.2 R_d$,
thus the inner bar cannot have its backbone
built out of orbits related to the x2 orbits in the outer bar. Further
work is needed to establish orbital support of double bars like
the ones in the model presented here.
The possible absence of an ILR makes the disc favourable to grow a bar through
the swing amplification of waves as it allows the feedback loop to 
complete \citep{Toomre1981}. On the other hand, since our bars form slowly 
in an initially rather stable disc (high Toomre's $Q$), and since they do not 
extend to their corotation radii, the mechanism proposed by 
\cite{Lynden-Bell1979} may play a role in their formation.
However, neither of these mechanisms 
anticipated formation of multiple bars.
\vspace{-0.4cm}
\section{Discussion and conclusions}
In this paper, we presented a model of a stellar disc, which 
spontaneously forms two bars that is markedly different from systems
simulated previously: the gravitational potential is dominated by the
dark halo (Fig.~\ref{fig:vc102}), the inner bar is large (Fig.~\ref{fig:denmapDB}), 
and the angular velocities of the bars are almost equal (Fig.~\ref{fig:pattspd}).

All numerical simulations of double bars to date assume gravity
dominated by stars in the region where the bars form. If Toomre's Q 
is low, then the outer bar forms rapidly, and an additional process 
is needed to induce the formation of the inner bar.
On the other hand, double bars can form spontaneously in pure N-body models 
when the disc is poorly coupled by its own self-gravity, which has to compete 
with the gravity of the massive halo or with thermal motions.
\cite{RautiainenSalo1999} were able to obtain double bars in their Model IV, 
in which they increased the Toomre parameter $Q$ in the central parts of the 
disc to $Q=3$, from $Q=1.5$ in the otherwise identical Model I, 
which returned a single bar only. In sets of cosmological N-body models by 
\cite{Curiretal2006}, double bars form when the disc-to-halo mass ratio is 
smallest in each set. Only single bars form in more massive discs in those models.
Our simulations presented in this paper confirm this trend, because they form 
double bars in the disc with high Q, which 
is dominated by dark matter halo. These findings 
indicate that the route to the formation of double bar may be different from 
that of a strong single bar, and the dark halo or hotter disc may play an important role.
Bar formation in our simulation scaled
to younger, smaller discs proceeds faster, but further work is needed
to study evolution of such spontaneously formed
double bars once the disc grows more massive.

In the majority of numerical models, both purely stellar and with 
a gaseous component, the pattern speed of the inner bar is 
significantly larger than that of the outer bar \citep{FriedliMartinet1993,
RautiainenSalo1999, DebattistaShen2007,Heller2007}.
Our model shows that an inner bar with the angular velocity
similar to that of the outer bar is also possible. Since throughout the 
evolution of our model, the two bars can be in any relative 
orientation (Fig.~\ref{fig:denmapDB}), the observed random orientation of the two 
bars \citep{ButaCrocker1993, FriedliMartinet1993} does not have 
to imply that pattern speeds of the bars differ significantly. 

The inner bar in our model is large -- it is about half
of the size of the outer bar for most part of the run. This is more
than the typical size ratio of the bars, being 0.12 \citep{ErwinSprake2002},
but size ratios up to 0.4 have been observed \cite[NGC 3358,][]{Erwin2004}. 
An inner bar supported by orbits inside the ILR of the outer bar 
cannot be too large \citep{MS2000}, but in our model the outer bar may have no
ILR, hence other orbits, without such size constraint, must support the inner 
bar here.

In summary, the model of double bars presented here indicates that
(1) formation of double bars may proceed under different 
conditions and in a different way than the formation of a single 
strong bar -- it may need dynamically hot stellar disc, possibly
dominated by the dark halo; 
(2) inner bars as large as half of the length of the
outer bar can last for a Hubble time or longer; 
(3) the difference 
between pattern speeds of the two bars can be minimal, yet the two 
bars can be observed in any relative orientation.
\vspace{-0.6cm}
\section*{Acknowledgement}
\noindent We would like to thank the anonymous referee for pointing out
that coupling of the disc by its self-gravity may play a role in 
formation of double bars, the remark that we incorporated in the 
revised text, and Peter Erwin for useful discussions.
K.S. acknowledges support from the Alexander von Humboldt 
Foundation. WM acknowledges the ESO Visiting Fellowship, which
allowed to initiate this project.

\end{document}